\documentclass[aip,jcp,reprint]{revtex4-1}
\usepackage{bm}
\usepackage{epsfig}
\usepackage{amsmath}
\usepackage{graphicx}
\usepackage{amssymb}
\usepackage[normalem]{ulem}
\usepackage[usenames]{color}
\usepackage{braket,calc}
\usepackage{amsfonts}
\usepackage{url}
\begin{document}

\def\tbar{\bar{t}_\text{c}}
\def\tf{t_\text{f}}
\def\ts{t_\text{s}}
\def\wden{w}
\newcommand{\vect}[1]{\vec{#1}}  
\def\fc{f_\text{c}}

\newcommand{\C}{\mathcal{C}}
\newcommand{\ee}{\mathrm{e}}
\newcommand{\eB}{\varepsilon_\mathrm{B}}
\newcommand{\ebt}{\tilde\varepsilon_\mathrm{B}}
\newcommand{\tobs}{t_\mathrm{obs}}
\newcommand{\nm}{n_\mathrm{max}}
\newcommand{\numb}{n_\mathrm{umb}}
\def\eS{\varepsilon_\text{S}}
\newcommand{\ns}{n_\text{s}}
\newcommand{\scL}{\mathcal{L}}
\newcommand{\eb}{\varepsilon_\mathrm{B}} 
\newcommand{\es}{\varepsilon_\mathrm{s}}
\newcommand{\rc}{r_\mathrm{c}}
\newcommand{\tc}{\theta_\mathrm{c}}
\newcommand{\pc}{\phi_\mathrm{c}}

\newcommand{\LJ}[1]{ \mathcal{L}_{#1} }
\newcommand{\ecp}{\varepsilon_\mathrm{cp}}
\newcommand{\ecc}{\varepsilon_\mathrm{cc}}

\newcommand{\sub}[1]{ _{\mathrm{#1}}}
\newcommand{\kbt}{k_\mathrm{B} T}
\newcommand{\kb}{k_\mathrm{B}}
\newcommand{\kt}{k_\mathrm{B}T}

\definecolor{Blue}{rgb}{0,0.0,1.0}
\definecolor{Red}{rgb}{1.0,0.0,0.0}
\definecolor{Green}{rgb}{0,1.0,0}
\newcommand{\mrpchange}[1]{\textcolor{Red}{#1}}
\newcommand{\mfhchange}[1]{\textcolor{Blue}{#1}}
\newcommand{\comment}[1]{\textcolor{Green}{#1}}
\newcommand{\changec}[2]{\textcolor{Red}{#1}\textcolor{Blue}{#2}}
\newcommand{\schange}[2]{{\sout{#1}}{\change{#2}}}


\title{Using Markov State Models to Study Self-Assembly}
\author{Matthew R. Perkett}
\author{Michael F. Hagan}
\email{hagan@brandeis.edu}
\affiliation{Martin Fisher School of Physics, Brandeis University, Waltham, MA, USA}

\begin{abstract}
Markov state models (MSMs) have been demonstrated to be a powerful method for computationally studying intramolecular processes such as protein folding and macromolecular conformational changes.    In this article, we present a new approach to construct MSMs that is applicable to modeling a broad class of multi-molecular assembly reactions.  Distinct structures formed during assembly are distinguished by their undirected graphs, which are defined by strong subunit interactions.  Spatial inhomogeneities of free subunits are accounted for using a recently developed Gaussian-based signature.  Simplifications to this state identification are also investigated. The feasibility of this approach is demonstrated on two different coarse-grained models for virus self-assembly.  We find good agreement between the dynamics predicted by the MSMs and long, unbiased simulations, and that the MSMs can reduce overall simulation time by orders of magnitude.
\end{abstract}

\maketitle

\section{Introduction}
The assembly of basic units into structures with increased size and complexity is central to biology, where examples of assembled structures include viruses (e.g. Refs.\cite{Caspar1962,Zlotnick2011,Speir2012,Hagan2014}), cell membranes, cytoskeletal filaments~\cite{Yang2010}, and ordered layers of proteins on bacterial surfaces ~\cite{Whitelam2010}. Assembly is also increasingly important to nanoscience, where interactions between colloidal particles are being engineered to drive assembly into sophisticated, functional materials~\cite{Sacanna2010} and DNA origami promises the ability to build structures of nearly limitless complexity  (e.g. Refs.~\cite{Rothemund2006,Sacca2012,Torring2011}).  An important focus of current research is understanding how the interactions between individual components determine assembly pathways, timescales, and fidelity for a target structure. Computational modeling can play a key role in determining assembly pathways and mechanisms, since most intermediates are transient and thus not readily characterized in experiments. However, simulating assembly is challenging because target structures can be orders of magnitude larger than their constituent components and assembly pathways typically surmount large free energy barriers, leading to timescales which greatly exceed computational limitations.

This paper is concerned with using Markov State Models (MSMs) to overcome the gap between assembly times and computationally accessible timescales. Many powerful enhanced sampling techniques have been developed to efficiently harvest computational trajectories that include barrier crossings or other rare events (e.g. \cite{Pan2008, Ovchinnikov2011,Bolhuis2002,Fischer2011,Elber2007,Pietrucci2009,Lei2004,Moroni2004,Dickson2010,Allen2005,Pfaendtner2009,Barducci2010,Zhang2007,Huber1996,Ferguson2011}).  However, many of these methods are limited by requiring \emph{a priori} knowledge of reaction coordinates, one or few pathways to completion, or only several metastable minima. In contrast, MSMs can be used to study assembly reactions characterized by multiple free energy barriers, a diverse ensemble of pathways, and pleomorphic products. Furthermore, MSMs are one of only a few methods \cite{Becker2012} that can describe non-stationary,  out-of-equilibrium dynamical processes.

While MSMs have been extensively developed in the context of protein folding\cite{Bowman2011,Bowman2010,Bowman2009,Swope2004,Swope2004a,Prinz2011,Park2006,Pande2010,Noe2009,Lane2011,Jayachandran2006,Chodera2011,Hinrichs2007,Chodera2007,Deuflhard2005}, existing approaches cannot describe the assembly of disconnected, permutable subunits. Here, we present a method to construct MSMs that is applicable to a wide variety of such assembly reactions. We test our approach on two models, which respectively describe the assembly of viral proteins around rigid nanoparticles and flexible polymers.

\section{Methods}
\subsection{Existing implementations of MSMs}
In this section we review how relatively short, unbiased simulations can be used to build an MSM and how this technique has been applied to study protein folding or conformational transitions.  The procedure begins by partitioning configurations from the short simulations into states such that conformations which interconvert rapidly are collected into the same state. The separation of timescales resulting from this partitioning ensures that the model is Markovian on timescales longer than a `lag time' $\tau$, meaning that the probability of transitioning to a new state only depends upon the current state.  Taking $\vect{P}(0)$ to be the vector of probabilities of being in each of the possible states of the system at time $t=0$, the state probability at time $t_\text{f}$ is given by
$\vect{P}(t_\text{f})=\bm{T}(\tau)^n \vect{P}(0)$
where $n=t_\text{f}/\tau$ and $\bm{T}(\tau)$ is the stochastic matrix of interstate transition probabilities estimated from the simulations at lag time $\tau$.

Determining a state decomposition that achieves the separation of timescales described above is a crucial aspect of building an MSM.  If the states do not sufficiently distinguish values of all of the slow degrees of freedom in the system, then the lag time $\tau$ at which the system becomes Markovian will be comparable to its longest relaxation timescale.  Since simulations must be greater than $\tau$ in length, the `short' simulations will approach the length of long, unbiased trajectories and thus the method will offer no computational savings.

Several approaches to determining state decompositions have been developed in the context of all-atom protein simulations\cite{Bowman2011, Bowman2010, Bowman2009, Swope2004, Swope2004a, Prinz2011, Park2006, Pande2010, Noe2009, Lane2011, Jayachandran2006, Chodera2011, Hinrichs2007, Chodera2007, Deuflhard2005, Sriraman2005, Chodera2006, Sorin2005, Elmer2005, Andrec2005,  Groot2001, Karpen1993, Singhal2004, Bowman2009a,  Beauchamp2011, Prinz2011a, Beauchamp2012, Noe2009, Voelz2010, Bowman2012, Perez-Hernandez2013, Schwantes2013}. In cases where a set of collective coordinates describing all of the slow degrees of freedom is known (or guessed) \emph{a priori}, biased sampling can be used to determine the free energy landscape as a function of these coordinates. States can then be defined based on local free energy minima (e.g. \cite{Swope2004,Sriraman2005,Chodera2006,Sorin2005,Elmer2005}).  Since it is rare to have \emph{a priori} knowledge of good collective coordinates, alternative approaches have been developed in which configurations are clustered based on geometric criteria, such as structural similarity (e.g. \cite{Andrec2005, Groot2001, Karpen1993, Singhal2004}).   Chodera et al. \cite{Chodera2007} developed an algorithm to refine an initial geometric partitioning of `microstates' into `macrostates' based on kinetics. Open source software packages such as MSMBuilder \cite{Bowman2009a, Beauchamp2011} and EMMA \cite{Senne2012} provide a suite of tools for building MSMs in this fashion and analyzing them.  This algorithm and similar approaches have been shown to be extremely powerful for the study of proteins, allowing for prediction of folding pathways and rates on even the supra-millisecond timescale (e.g. Refs. \cite{Beauchamp2011,Prinz2011a,Beauchamp2012,Noe2009,Voelz2010}) as well as identifying hidden allosteric sites \cite{Bowman2012}.  Recently, systematic approaches to find optimal coordinates for constructing MSMs have been developed \cite{Perez-Hernandez2013,Schwantes2013}.

\subsection{Building MSMs for assembly systems}
\label{sec_build_msm}
In contrast to protein folding, where each residue in the protein has a unique index, assembly subunits are permutable and thus cannot be indexed.  Therefore, existing algorithms for determining state decompositions cannot be directly applied to self-assembly. Here, we describe several approaches to creating a state decomposition based on the network of subunit interactions and their positions relative to heterogeneous nucleation sites.

We consider systems in which subunits can assemble either through homogeneous nucleation to form a single component structure (e.g. an empty virus capsid shell) or through heterogeneous nucleation around a scaffold (e.g. a polymer or a nanoparticle) to form a multicomponent structure. To simplify the presentation, we consider one type of subunit  and focus on only the largest assembled structure in the system at any given time, but the approach can be generalized to multiple subunit species and assemblies.  Our approach can be applied to systems which assemble via reversible or irreversible interactions; in both cases we will describe a strong interaction as a `bond'.


To generate a state decomposition, we categorize subunits into three classes: \textbf{class I}: bonded subunits in the assemblage, \textbf{class II}: subunits bonded to the scaffold, but not the assemblage, and \textbf{class III}: unbonded, free subunits.  In the case of homogeneous nucleation, there are no class II subunits (since there is no scaffold), and class III subunits can be ignored if subunit association to the scaffold is sufficiently reaction-rate limited that the density of free subunits is spatially uniform.

For systems that assemble into well-defined structures, fluctuations in bond distances and angles are fast compared to bond formation and breakage. By averaging over these short timescale fluctuations, the unique structure of a growing cluster can be defined by the class I subunit bonding network.  More precisely, each cluster is converted into an undirected graph with nodes corresponding to subunits and edges corresponding to bonds between the subunits (see Fig. \ref{fig-graphIso-explanation} in Appendix \ref{appendix_graphIso}).  This ensures a consistent state definition that is unaffected by exchanges of subunits with different indices, short-timescale conformational fluctuations, or rigid body motions of the assemblage. Class I subunits can be further sub-partitioned by including the distance from the scaffold to each subunit, but this was unnecessary for the systems that we considered.  Class II subunits can be handled in a similar manner by considering the subunit-scaffold bonding network, but we found that it was only necessary to track the total number of class II subunits.

Class III subunits must be included in the state decomposition when their association to the cluster or scaffold approaches the diffusion limit, which results in  density inhomogeneities.  Free subunit positions fluctuate rapidly since they are not involved in strong interactions, and thus we use their density distribution rather than their positions to decompose states. We follow the approach developed by Gu et al. \cite{Gu2013} to include solvent degrees of freedom in protein folding MSMs.  We define a vector $\bm{\wden}$, in which each index corresponds to a subunit in the cluster or a residue of the scaffold and each component corresponds to a distance-weighted density of free subunits around the indexed subunit:
\begin{equation}
\wden_i \equiv \sum_{j \in \text{free subunits}} e^{-|\vec{r}_j-\vec{r}_i|^2/\sigma_\text{d}}
\label{eq_weighted_density}
\end{equation}
with $\vec{r}_i$ and $\vec{r}_j$ as the positions of cluster/scaffold subunit $i$ and the free subunit $j$.  $\sigma_\text{d}$ is an adjustable decay length that sets the scale for relevent interactions between free subunits and the scaffold or growing cluster.  This definition weighs nearby subunits more heavily since they are more likely to associate with the cluster or scaffold.  Whether cluster subunits, scaffold residues, or both need to be considered in this definition depends on which association reactions approach the diffusion limit. For example, in the case considered below where subunit adsorption onto the nanoparticle approaches the diffusion limit but subunit-subunit associations do not, it is only necessary to include the nanoparticle in $\bm{\wden}$.

Finally, if scaffold internal degrees of freedom (e.g. conformations of a polymer) evolve slowly, these should also be included in the state definition. Because the scaffold residues (i.e. polymer segments) are indexable, these degrees of freedom can be treated via the existing RMSD-based approach (e.g. in MSMBuilder \cite{Beauchamp2011,Bowman2009a} and EMMA \cite{Senne2012}).

{\bf Reducing the number of states.}
As the size of the target structure increases, the number of distinct assembly intermediates and hence the number of unique graphs grows rapidly.  The number of states could become intractable if class II or class III subunits were included in the state definition.  There are two routes to reduce the number of states.  First, kinetic data can be used to group microstates that interconvert rapidly into macrostates by following the approach based on Perron cluster analysis in MSMBuilder \cite{Deuflhard2005,Beauchamp2011,Bowman2009a} and EMMA \cite{Senne2012}.  Second, \emph{a priori} knowledge of the system can be used to reduce the number of unique states.  Since assembly into a target structure with high fidelity generally requires weak subunit-subunit interactions \cite{Hagan2014,Grant2011,Zlotnick2003}, subunits in a cluster with only one bond rapidly dissociate.  Thus, the edges corresponding to these interactions can generally be neglected when building graphs.  For the models considered here, we found that fluctuations about the most compact, highly bonded structures are rapid enough that it was sufficient to consider only the number of subunits in a cluster. Several alternative simplified descriptions are discussed in Appendix \ref{appendix_graphIso}.

\subsection{Generating The Transition Matrix}
\label{sec_trans_mat}
The transition probability matrix $\mathbf{T}(\tau)$ is calculated by column-normalizing the count matrix $\mathbf{C}(\tau)$, in which each element $C_{ji}$ gives the total number of transitions from state $i$ to state $j$ measured at a lag time $\tau$.  The count matrix can be calculated from many, relatively short, unbiased trajectories run in parallel.  Because of the Markov property, the initial conditions for these trajectories can be chosen to efficiently generate good statistics for all of the relevant transition elements.

When no information is available \emph{a priori} about which transition elements are most significant, one can use a ratcheting procedure (Appendix~\ref{sec:protocol}). Many simulations are run in parallel for a time $\ts$, which must be longer than the lag time $\tau$ but can be much shorter than the longest relaxation timescale.  Microstates are then determined from coordinates saved during these trajectories, and a new ensemble of trajectories is started with initial conditions preferentially chosen from the microstates with the poorest sampling. This procedure is repeated until $\mathbf{T}$ has satisfactorily converged.  Once sufficient statistics have been gathered to crudely estimate $\mathbf{T}$, more systematic adaptive sampling \cite{Bowman2010} can be used to choose initial conditions that will reduce the statistical uncertainty of the MSM.  However, the initial ratcheting procedure already allows for tremendous speed up in comparison to long, brute force simulations as it enables the system to cross free energy barriers in linear rather than exponential simulation times.  Note that because the protocol does not generate initial conditions according to the equilibrium distribution, the count matrix $\mathbf{C}$ should not be symmeterized when simulating assembly dynamics.  In fact, even if $\mathbf{C}$ is estimated from long, unbiased trajectories that achieve formation of the target structure it should not be symmeterized when calculating dynamics, since assembly to the target structure is an out-of-equilibrium process.

\subsection{Analysis of MSMs}
\label{sec_msm_analysis}
In this subsection we briefly review analysis of constructed MSMs and discuss an application which is useful for  analyzing assembly reactions.
Upon spectral decomposition of the transition matrix, the time-dependent state probabilities can be written as
\begin{align}
\vec{P}(t;\tau) & = \sum_{i=1}^{N} \ket{i} \bra{i} \ket{\vec{P}(0)} e^{-\lambda_i t} \nonumber \\
\lambda_i &= -\log(\omega_i)/\tau
\label{eq_state_prob_2}
\end{align}
where $\omega_i$ is the $i^{\text{th}}$ eigenvalue of $\mathbf{T}(\tau)$ and $\bra{i}$ and $\ket{i}$ are the corresponding left/right eigenvectors, which are assumed to be normalized.  Since $\mathbf{T}(\tau)$ is generally not Hermitian the left and right eigenvectors are not equivalent.  Because the rate matrix is stochastic, there is only one unit eigenvalue, whose associated right eigenvector corresponds to the equilibrium distribution, while all other eigenvalues are positive and real \cite{Swope2004a}.   The implied timescale, $|\lambda_i|^{-1}$, corresponds to the relaxation timescale for  eigenmode $i$.  For lag times on which the system satisfies the Markov assumption, the calculated implied timescales are nearly independent of $\tau$ \cite{Prinz2011a}.  Checking the convergence of the implied timescales is useful for selecting an appropriate $\tau$, but does not guarantee that the model is Markovian, which also requires converged eigenvectors \cite{Prinz2011a}.

{\bf Self-assembly reactions.}
It is often useful to calculate the completion fraction $\fc(t)$, which is defined as the fraction of structures in the target state as a function of time.  This quantity can be compared to light scattering or size exclusion chromatography experiments \cite{Prevelige1993, Zlotnick1999, Zlotnick2000, Casini2004, Chen2008, Berthet-Colominas1987,Kler2012,Tsiang2012,Kumar2010}.  With  $\vect{P}$ ordered such that index 1 corresponds to the initial, unassembled state and the largest index $N$ corresponds to the target state, $\fc$ is given by $P_N(t)$.  Inserting  $P_i(0)=\delta_{i,1}$, with $\delta$ the Kroniker delta, into   Eq.~\eqref{eq_state_prob_2} gives
\begin{equation}
f_{c}(t) = \sum_{i=1}^{N}\ket{i}_{N} \bra{i}_{1} e^{-\lambda_i t}
\label{eq_frac_complete}
\end{equation}
where $\bra{i}_{n}$ indicates the $n^{\text{th}}$ index of $\bra{i}$.  The mean completion time $\tbar = \int_{0}^{\infty} \left(\frac{d}{dt} \fc(t)\right) t dt$ then follows as
\begin{equation}
\tbar = \sum_{i=2}^{N}\ket{i}_{N} \bra{i}_{1} \frac{1}{\lambda_i} e^{-\lambda_i t}
\label{eq_mean_comp_time}.
\end{equation}

\subsection{Analysis using transiton path theory \cite{E2010, Metzner2006,Noe2008, Noe2009, Voelz2010, Metzner2009}} \label{sec:methods_tpt}
Insight into assembly mechanisms can be obtained from MSMs using Transition Path Theory (TPT) \cite{E2010, Metzner2006}, which has been developed in the context of MSMs in Refs \cite{Noe2008, Noe2009, Voelz2010, Metzner2009}.  We state two of these results which are particularly useful for analyzing assembly reactions here.  The microstates that correspond to the transition state ensemble can be identified by calculating the committor probability  for each state, which is the probability that a dynamical trajectory initiated from a given state will subsequently visit the target state\cite{Dellago2002,Prinz2011b}.  We define $A$ as the set of mostly unassembled states that rapidly interconvert (the reactant states), $B$ as the target structure (the product state), and $I$ as all other states (the intermediate states). The forward committor probability, $q_i^+$, is the probability that a trajectory started in state $i$ will visit $B$ before $A$ and is given by solving \cite{Noe2009}
\begin{equation}
q_i^+ - \sum_{j \in I} T_{ij} q_j^+ = \sum_{j \in B} T_{ij}
\label{eq_committor}.
\end{equation}
Similarly, the backward committor probability $q_i^-$ is the probability that the system was more recently in state $A$ than $B$.  For an equilibrium system, the committor probabilities are related by \cite{Noe2009} $q_i^-=1-q_i^+$, but for the models considered here, the target structure acts as an absorbing state (see Appendix~\ref{sec:protocol}), which gives $q_i^-=1$ for $i\notin B$.

The relative probabilities of different assembly pathways can be calculated from the flux between states, which is given by \cite{Noe2009} $f_{ij} = \pi_i q_i^- T_{ij} q_j^+$, with $\pi_i$ as the stationary probability of being found in state $i$.  Since $f_{ij}$ contains non-productive loops that are not on the pathway to completion, the forward flux $f_{ij}^+$ is defined by subtracting out these contributions \cite{Noe2009}:
\begin{equation}
f_{ij}^+ = \text{Max}(0,f_{ij} - f_{ji})
\label{eq_forward_flux}
\end{equation}

\section{Models}
To test and benchmark our MSM framework, we consider two previously studied models for viral assembly, which differ in their level of detail and more importantly in the type of cargo being packaged. Both models represent capsid protein subunits as rigid bodies with excluded volume geometries and orientation-dependent interactions, designed such that the lowest energy structure is an icosahedron with 20 subunits. Each subunit can be thought of as describing a trimer of proteins that form a $T{=}1$ capsid \cite{Hagan2014}.

\begin{figure}[hbt]
  \begin{center}
  \includegraphics[width=\columnwidth]{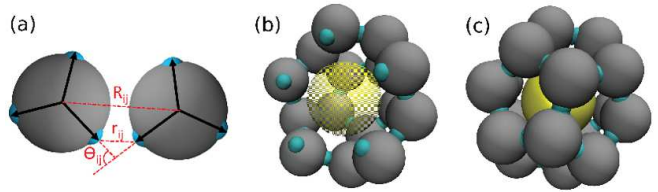}

  \caption{Patchy sphere model geometry. {\bf (a)} Geometry of two attracting subunits with the bond vectors depicted as arrows and attractors colored teal.  The angle between each of the subunit bond vectors is $108^{\mathrm{o}}$, and the interactions are described in the Appendix (Eqs. \eqref{eq_potential}).  The dihedral angle $\phi_{ij}^b$ is not shown. {\bf (b)} A cutaway view of the complete capsid showing the nanoparticle.  {\bf (c)} The complete capsid, which contains 20 subunits arranged with icosahedral symmetry.}
  \label{fig_patchy_model}
  \end{center}
\end{figure}

{\bf Patchy Sphere Model.}
The first model is motivated by experiments in which capsid proteins assemble around nanoparticles functionalized with negative charge \cite{Tsvetkova2012,Sun2007,Huang2007,Dixit2006,Daniel2010,Chen2005,Chen2006,Hagan2008,Hagan2009,Elrad2008,Siber2010,He2013}.  The subunit excluded volume is spherically symmetric, and three attractive patches (bond vectors) are rigidly fixed to the subunit, with each pair of bond vectors forming an angle of $108^\circ$ (see Fig. \ref{fig_patchy_model} and Eq.~\eqref{eq_potential}).  There is a favorable interaction between subunits when (1) the ends of bond vectors nearly overlap, (2) the bond vectors are nearly anti-parallel, and (3) the secondary bond vectors are nearly coplanar.  Twenty subunits realizing these conditions results in the minimum energy target structure (a complete capsid) shown in Fig. \ref{fig_patchy_model}. The interaction strength is tuned by the parameter $\eB$. The nanoparticle has a spherical excluded volume and short-range attractive interactions with capsid subunits, which are tuned by the parameter $\eS$ and qualitatively represent screened electrostatic attractions. More details about this model can be found in Appendix \ref{sec:patchyAppendix}.

\begin{figure}[hbt]
  \begin{center}
  \includegraphics[width=\columnwidth]{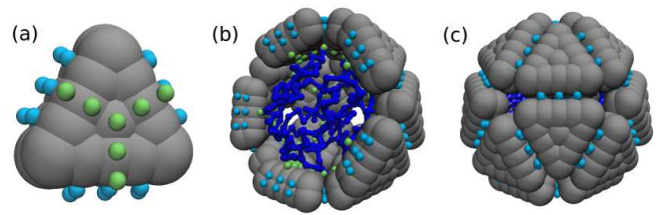}
  \caption{Triangle model geometry. {\bf (a)} Subunit geometry with grey excluders, green polymer attractors, and teal subunit attractors.  Subunits experience an attractive interaction when the subunit attractors nearly overlap. {\bf (b)} A cutaway view of the complete capsid with the encapsulated polymer shown in blue.  {\bf (c)} The complete capsid, which contains 20 subunits.}
  \label{fig_triangle_model}
  \end{center}
\end{figure}

{\bf Triangles Model.}
This model represents a capsid protein subunit using multiple, spherical `excluders' that enforce excluded volume and spherical `attractors' with short-range, pairwise attractions tuned by the parameter $\ecc$.  Excluders and attractors are arranged so that the minimum energy capsid is an icosahedron (Fig. \ref{fig_triangle_model}).  The cargo considered here is a self-avoiding bead-spring polymer with a persistence length comparable to that of single-stranded RNA (ssRNA) with no base pairing.  Polymer beads experience short range attractive interactions with polymer-attractors located on the bottom of protein subunits (see Fig. \ref{fig_triangle_model}); the interaction strength is tuned by the parameter $\ecp$.  This model has previously been used to study assembly around a polymer\cite{Elrad2010} and empty capsid assembly\cite{Hagan2011}.  Similar models were applied to empty capsids by Rapaport  \cite{Rapaport1999, Rapaport2004, Rapaport2008} and Nguyen et al. \cite{Nguyen2007}. More details are given in Appendix \ref{sec:trianglesAppendix}.

{\bf Simulations and units.} Subunit positions and orientations are propagated using overdamped Brownian dynamics according to a second order predictor-corrector algorithm~\cite{Branka1999,*Heyes2000}. To represent an experiment with excess capsid protein, each simulation includes a single scaffold (nanoparticle or polymer) and is coupled to a bulk solution by performing grand canonical Monte Carlo moves, in which subunits at the periphery of the simulation box are exchanged with a reservoir at fixed chemical potential with a frequency consistent with the diffusion limited rate \cite{Hagan2008,Elrad2010}. To obtain dimensionless units, we rescale energies by $\kt$ and times by a characteristic diffusion timescale (see Appendix \ref{sec:modelsAppendix}).

\section{Results and Discussion}
\begin{figure}[hbt]
  \begin{center}
  \includegraphics[width=\columnwidth]{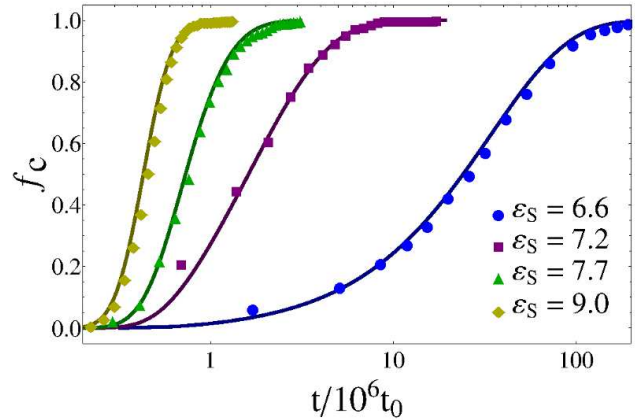}
  \caption{Patchy sphere model: Fraction complete ($\fc$) as a function of time from the brute-force calculations (symbols) and the MSM calculations (lines).  The subunit-subunit binding energy parameter is $\eB=10$ for all simulations and the values of the subunit-nanoparticle interaction parameter $\eS$ are indicated on the plot.  MSM states are defined by the number of subunits adsorbed to the nanoparticle and the largest cluster size, except for $\eS{=}9$, which also includes Eq. \eqref{eq_weighted_density} for density variations around the nanoparticle with $\sigma_\text{d}=4\sigma$.  Notice that a semi-log scale is used to accomodate a wide range of assembly times. The brute force estimates of $\fc(t)$ are calculated from the completion times for $500$-$1,000$ unbiased simulations.}
  \label{fig-patchy-fc}
  \end{center}
\end{figure}

\begin{figure}[hbt]
  \begin{center}
  \includegraphics[width=\columnwidth]{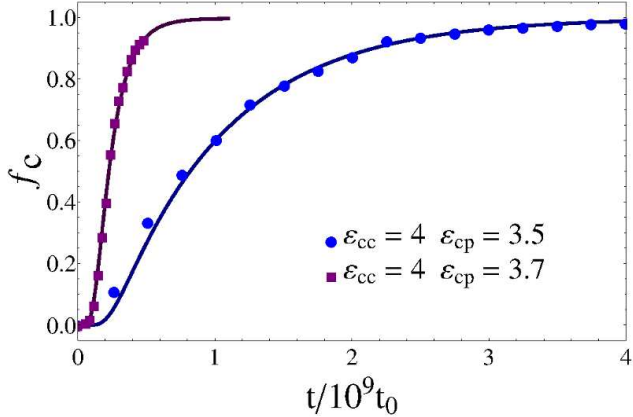}
  \caption{Triangles model: Fraction complete as a function of time for the brute-force calculations (symbols) and the MSM calculations (lines) for indicated values of  the subunit-subunit ($\ecc$) and subunit-polymer ($\ecp$) interaction parameters.  MSM states are defined by the number of subunits adsorbed to the polymer and the largest cluster size.}
  \label{fig-triangles-fc}
  \end{center}
\end{figure}

We performed simulations over a wide range of parameters to test the ability of the MSMs to accurately reproduce assembly dynamics and to determine the extent of computational speed up in comparison to brute force calculations.  To evaluate accuracy, we compare the MSM (Eq. \eqref{eq_frac_complete}) and brute force dynamics predictions for the cumulative distribution of assembly times, $\fc(t)$ in Figs. \ref{fig-patchy-fc} and \ref{fig-triangles-fc}.   We find that this comparison provides a stringent test of the MSM, as it requires an accurate estimate of all statistically relevant elements of the transition matrix.  In particular, capturing the assembly lag phase (the time before the first target structures appear, see Fig. \ref{fig-patchy-fc}) requires accurate estimates for as many as 20 implied timescales and their associated eigenvectors.

\begin{figure}[hbt]
  \begin{center}
  \includegraphics[width=\columnwidth]{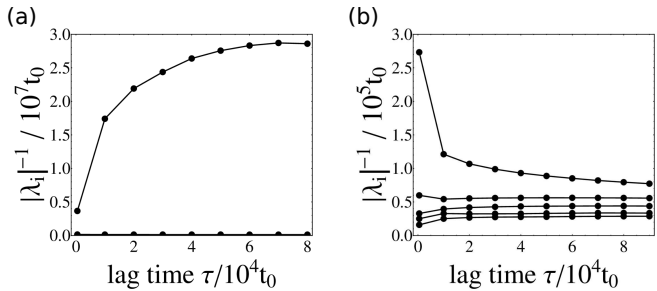}
  \caption{Implied timescales $|\lambda_i|^{-1}$ for the patchy sphere model.  The five largest eigenvalues are shown for {\bf (a)} slow, nucleation dominated assembly ($\eB = 10$, $\eS = 6.6$) and {\bf (b)} rapid assembly ($\eB = 10$, $\eS = 9$).  For (a), the largest implied timescale (excluding the unit eigenvalue) corresponds to the nucleation timescale.}
  \label{fig_implied}
  \end{center}
\end{figure}

The assembly time distributions are accurately predicted from the MSMs for a wide range of parameter values that represent different assembly mechanisms (Figs. \ref{fig-patchy-fc} and \ref{fig-triangles-fc}).  In the patchy sphere model, assembly for the weakest subunit-nanoparticle interaction ($\eS{=}6.6$) is heavily nucleation-dominated, while for the strongest value ($\eS{=}9$) the nucleation timescale is comparable to the elongation timescale \cite{endres2002} (i.e. the time required for a critical nucleus to grow to completion \cite{Hagan2010}).  These two scenarios are distinguished by the spectrum of implied timescales (Fig. \ref{fig_implied}). MSMs corresponding to nucleation-dominated parameter sets are characterized by a wide separation between the two largest implied timescales, whereas the MSM corresponding to $\eS=9$ yields a dense spectrum of implied timescales.  As discussed in section \ref{subsec_tpt_analysis}, the different parameter values give rise to very different assembly pathways as well.

For most parameter sets, we found that the minimal state definition capable of reproducing assembly dynamics included the number of subunits in the largest cluster and the number of subunits adsorbed to the nanoparticle or polymer.  However, it is worth noting that the ratcheting procedure used to estimate the transition matrix considered not only the cluster size, but also the number of intra-cluster bonds (Appendix~\ref{sec:protocol}).  Ratcheting was less efficient when only the cluster size was considered, which suggests that the bonding network within a cluster is important. For the parameter set with the strongest subunit-nanoparticle interaction strength ($\es{=}9$ in Fig.~\ref{fig-patchy-fc}), subunits rapidly adsorbed onto the nanoparticle and it was also necessary to include the free subunit density distribution (Eq. \eqref{eq_weighted_density}) when building the MSM.

Although a simple state definition yields accurate results for these models, a more detailed order parameter that describes the cluster structure will be required in other situations.  To show that MSM construction is feasible even with our most general definition, we also generated MSMs using the graph coordinate (described in the Appendix~\ref{appendix_graphIso}).  As shown in Fig. \ref{fig-graphIsoVsnCluster} the predicted assembly time distributions are identical to those predicted from the simpler coordinates.\footnote{Convergence required a $50\%$ increase in total simulation time because statistics were reduced by the increased number of states ($18,000$ states with the graph compared to $220$ with the simple state definition). However, this increase could be at least partially eliminated through lumping of microstates.}

\begin{figure}[bt]
  \begin{center}
  \includegraphics[width=\columnwidth]{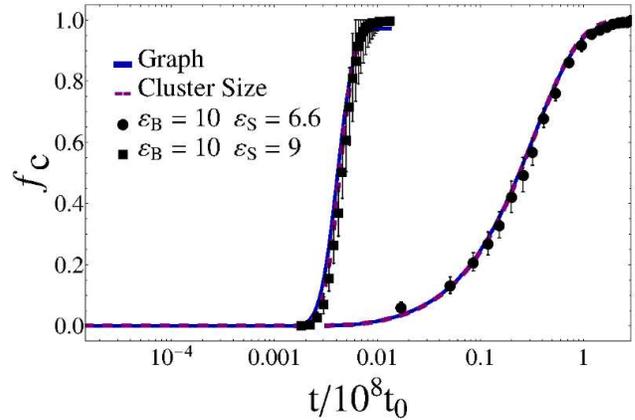}
  \caption{Feasibility of the graph coordinate.  The fraction complete for the patchy sphere model calculated from MSMs defined using the number of subunits adsorbed to the nanoparticle and either the largest cluster size (dashed line) or the graph coordinate (solid line) for indicated parameter values.  Data from long, unbiased simulations is shown as points.}
  \label{fig-graphIsoVsnCluster}
  \end{center}
\end{figure}

\begin{figure}[hbt]
  \begin{center}
  \includegraphics[width=\columnwidth]{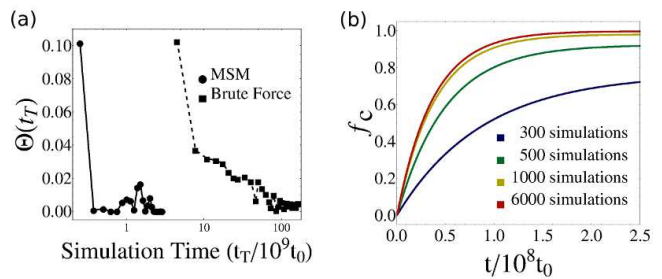}
  \caption{MSM convergence data for patchy spheres ($\eB = 10$, $\eS = 6.6$).  {\bf (a)} The scaled error (Eq.~\eqref{eq_convergence}) of the mean completion time $\bar{t}_c$ as a function of total simulation time $t_\mathrm{T}$ for the MSM calculation and straightforward dynamics. {\bf (b)} Convergence of the fraction complete $\fc$ as more trajectories are used to build the MSM.  The mean completion time is about 100 times the length of the short simulations used for this parameter set.  The MSM has converged after about $1,000$ short simulations, which corresponds to the total CPU time of 10 brute force simulations. Without the MSM, hundreds of brute force simulations are required to achieve comparable precision.}
  \label{fig_convergence}
  \end{center}
\end{figure}

{\bf Testing convergence.}   To test that a system is Markovian on timescales corresponding to the lag time $\tau$ used to build the transition matrix, one typically checks that the implied timescales are nearly lag-time-independent for lag times equal to or exceeding $\tau$ (Fig.~\ref{fig_implied}). For assembly systems, convergence can be more stringently tested by determining if the predicted assembly time distribution $\fc$, which depends on all of the implied timescales and associated eigenvectors, becomes independent of lag time (Fig.~\ref{fig_convergence}).

{\bf Simulation Time.}
To assess the computational speed up afforded by MSMs for our systems, we calculated a scaled error $\Theta$ for the estimated mean assembly time $\bar{t}_c$:
\begin{equation}
\Theta(t_\mathrm{T}) = \left(\frac{\bar{t}_c(t_\mathrm{T}) - \bar{t}_c(\tf)}{\bar{t}_c(\tf)}\right)^2
\label{eq_convergence}.
\end{equation}
Here $t_\mathrm{T}$ is the total simulation time accrued during the short trajectories used to estimate the transition probability matrix; $t_\mathrm{T}$ was varied by changing the number of short trajectories.  We neglected computational overhead associated with initializing short trajectories and spectral decomposition of the transition matrix, as these factors were negligible. We also calculated $\Theta$ as a function of simulation time for straightforward dynamics, with $t_\mathrm{T}$ varied by changing the number of trajectories used to estimate $\bar{t}_c(\tf)$. As shown in Fig \ref{fig_convergence}, the MSM calculation converges with an order of magnitude less simulation time than the estimate based on straightforward dynamics.  As expected, the magnitude of speedup depends on the separation of timescales, with greater speed up for large nucleation barriers and only limited speed up in growth-dominated regimes with dense timescale spectra.  Importantly, the method is not hindered by multiple, large nucleation barriers such as tend to occur for low values of $\eB$ in these capsid assembly systems \cite{Hagan2006}.

\subsection{Analysis using Transiton Path Theory.}
\label{subsec_tpt_analysis}
\begin{figure}[hbt]
  \begin{center}
  \includegraphics[width=\columnwidth]{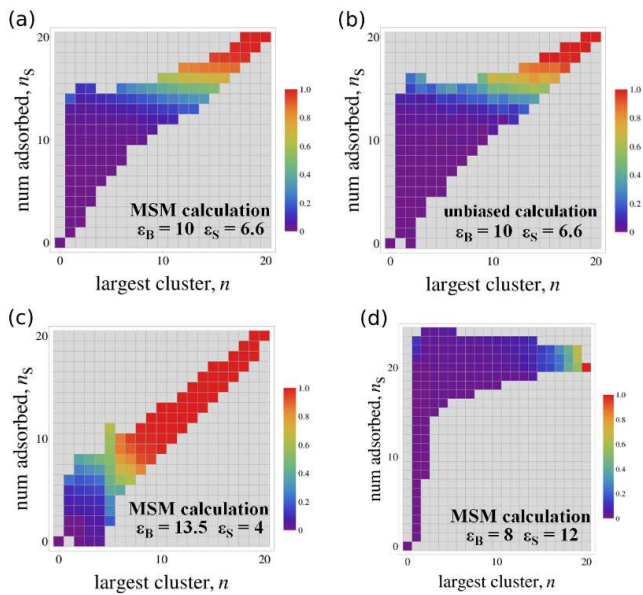}
  \caption{Patchy sphere committor probabilities (Eq. \eqref{eq_committor}). The top row compares committor probabilities calculated using {\bf (a)} the MSM and {\bf (b)} brute force dynamics for the same parameter set, while {\bf (c)} and {\bf (d)} show committor probabilities calculated using the MSM for two parameter sets for which assembly proceeds by different pathways. Parameter values are indicated on each plot.}
  \label{fig_committor}
  \end{center}
\end{figure}

As a further test of the MSM approach and its ability to provide mechanistic insight, we applied Transition Path Theory (TPT) \cite{Noe2009} (see section \ref{sec:methods_tpt}) to MSMs for several patchy sphere parameter sets.  First, we calculated committor probabilities as functions of the cluster size $n$ and the number of adsorbed subunits $\ns$. Committors calculated using the MSM (Fig. \ref{fig_committor}a) closely agree with those calculated from straightforward dynamical trajectories (Fig. \ref{fig_committor}b).  Provided that the chosen collective variables are suitable reaction coordinates \cite{Dellago2002}, states with committor probabilities near $q_i=1/2$ correspond to critical nuclei, or coordinates from which complete assembly or complete disassembly are equally probable. Fig. \ref{fig_committor} reveals that the ensemble of critical nuclei depends sensitively on parameter values.  For instance, with moderate strengths of subunit-subunit and subunit-nanoparticle interactions (Fig. \ref{fig_committor}a) critical nuclei occur for a narrow range of adsorbed subunits $\ns\approx15$ but include a broad range of $n=10-15$ subunits in the cluster, indicating that subunit adsorption is the controlling degree of freedom, while the assembly state of the adsorbed subunits is fluctuating rapidly during nucleation.  In contrast, for weak subunit-subunit and strong subunit-nanoparticle interactions (Fig. \ref{fig_committor}d), for which subunit adsorption is rapid but assembly requires a rare fluctuation to a large cluster size, the critical nuclei include a narrow range of $n$.

\begin{figure}[hbt]
  \begin{center}
  \includegraphics[width=\columnwidth]{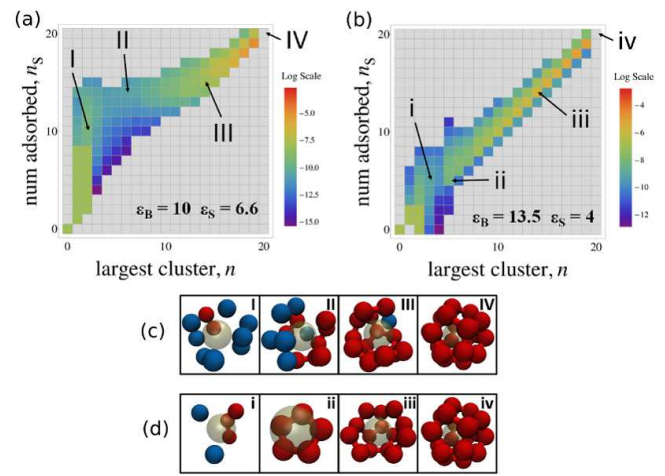}
  \caption{{\bf (a)},{\bf (b)}: The total forward flux (Eq. \eqref{eq_forward_flux}) through each state as calculated from the MSM for patchy spheres, with indicated parameter values.  Note that values of the forward flux are colored according to a log scale.  {\bf (c),\bf (d):} snapshots depicting the typical assembly pathways for the parameter sets in (a) and (b) respectively.  The Roman numeral labels indicate points on the flux diagrams to which the snapshots correspond. Bonded subunits are colored red, unbonded subunits adsorbed to the nanoparticle are colored blue, and the nanoparticle is rendered translucent.}
  \label{fig_flux}
  \end{center}
\end{figure}

To determine how parameter values influence assembly mechanisms, we used Eq.~\eqref{eq_forward_flux} to calculate the total forward flux through each Markov state (Fig. \ref{fig_flux}).  The sequence of states with the largest flux connecting the initial and target configurations corresponds to the most probable assembly pathway. We see that for nucleation dominated parameters ($\eB{=}10$, $\eS{=}6.6$, Fig. \ref{fig_flux}a,c) on average 15 of the 20 subunits required to form a capsid have adsorbed onto the nanoparticle in a disordered manner before any significant assembly occurs, indicating that assembly proceeds according to the \emph{en masse} mechanism defined in Refs \cite{Hagan2008,Elrad2010,Garmann2013}.  For weaker subunit-nanoparticle interactions and stronger subunit-subunit interactions ($\eS{=}4$, $\eB{=}13.5$, Fig. \ref{fig_flux}b,d), the predominant assembly pathway changes to a nucleation and growth scenario in which the number of adsorbed subunits and the cluster size proceed in lockstep, because adsorption of unassembled subunits is only transient.  The calculated assembly pathways closely agree with those determined from brute force dynamics.

\section{Discussion and Conclusions}
In this paper, we propose a general approach to construct MSMs for assembly systems, in which undirected graphs categorize the interaction network of growing clusters.  We also discuss several simplified descriptors of the assembly state. A Gaussian-based signature \cite{Gu2013} can be used to describe the density distribution of free subunits; however, this parameter can be neglected when subunit association is reaction-rate limited. We find that MSMs constructed from trajectories that are orders of magnitude shorter than the mean assembly time accurately describe dynamics on all assembly timescales for two different model systems.  Analysis of the constructed MSMs using transition path theory \cite{Noe2009} reveals predominant assembly pathways and the transition state ensembles as functions of control parameter values.  This information can be used to infer mechanisms and to identify key intermediates that could be targeted to effect changes in assembly yields or structures.

A key requirement of the MSM approach is determining a state decomposition for which the system is Markovian on a timescale short in comparison to the assembly time. Based on the success of our MSM calculation for two different models, we expect our approach to state decomposition to apply to a wide variety of assembly reactions.  Our structure-based coordinates are particularly well-suited for reactions in which subunit association and dissociation timescales are slow in comparison to the local relaxation of a cluster (i.e. fluctuations around preferred interaction distances and angles).  Although our simple ratcheting procedure for estimating the transition probability matrix produced significant speedup, convergence times could be further improved by refining state definitions based on interconversion kinetics  \cite{Bowman2010}.

In the two systems considered here, we found that state decompositions based on only the size of the largest cluster and the number of subunits adsorbed to the scaffold were sufficient for describing the dynamics for most parameter sets.  Generalizing the state definition to account for other partially assembled clusters did not improve the accuracy or convergence of the MSM.  However, this generalization could be important for systems in which there is an ensemble of large, partially assembled intermediates.  This situation occurs when free energy barriers to nucleation are small, allowing for the rapid and spontaneous formation of clusters.  For example, many capsid assembly simulations  have focused on parameter sets where nucleation and elongation timescales are comparable, since assembly is most readily accessible by straightforward dynamics at these parameters \cite{Hagan2014}.  However, this regime lacks the large separation of timescales that MSMs (and other enhanced sampling techniques) exploit to the greatest advantage.

The largest speed up  will be achieved in systems with one or more large free energy barriers.  Under these conditions nucleation is a true rare event and considering only the largest cluster is likely sufficient for state decomposition (see Ref.\cite{Ouldridge2010} for further discussion).  Such parameter sets are also most relevant to most experiments, where nucleation is rate limiting and concentrations of intermediates are often so low as to be undetectable (e.g. \cite{Hagan2014,Zlotnick1999}).  The MSM approach described here enables simulating assembly at these experimentally relevant parameter values.

\begin{acknowledgments}
We thank Vijay Pande for enlightening discussions and for suggesting to develop a coordinate based on subunit densities. This work was supported by Award Number R01GM108021 from the National Institute Of General Medical Sciences. We also acknowledge support by NSF-MRSEC-0820492.  Computational resources were provided by the National Science Foundation through XSEDE computing resources (Open Science Grid and Trestles) and the Brandeis HPCC.
\end{acknowledgments}

\appendix
\section{Model details}
\label{sec:modelsAppendix}
\subsection{Patchy sphere model}
\label{sec:patchyAppendix}
In this model, adapted from Ref.~\cite{Hagan2008}, the minimum energy structure is a complete capsid of 20 subunits encapsulating the nanoparticle.  Subunits have a spherical excluded volume with three attractive patches, or bond vectors, that are separated by $108^\circ$ and rotate rigidly with the subunit.  The attractive interaction between two complementary bond vectors on respective subunits $i$ and $j$ is maximized when (1) the distance between the attractors $r_{ij}^b$ is minimized, (2) the angle $\theta_{ij}^b$ between bond vectors is minimized, and (3) the dihedral angle $\phi_{ij}^b$ calculated from two secondary bond vectors, which are not involved in the primary interaction, is minimized.  The schematic of subunit interactions is shown in Fig. \ref{fig_patchy_model}.  Minimizing $\phi_{ij}^b$ creates an interaction that resists torsion and enforces angular specificity commensurate with a complete capsid.  The potentials are given by Eqs \eqref{eq_potential}  \cite{Hagan2008} \begin{eqnarray}
U &=& U_\text{rep}(R_{ij}) + \sum_b U_\text{att}(r_{ij}^b)S(\theta_{ij}^b, \phi_{ij}^b) \nonumber \\
U_\text{rep}(R_{ij}) &=&  \LJ{12}(R_{ij},2^\frac16 \sigma,\sigma)  \nonumber \\
U_\text{att}(r_{ij}) &=&  \eb \LJ{12}( (r_{ij} + 2^\frac16 \sigma), 2^\frac16 \sigma, \rc ) \nonumber \\
S(\theta,\phi) &=& \frac14 \ \Theta(\theta - \theta_c) \Theta(\phi - \phi_c)  \nonumber \\
        &\phantom{=}&\left( \cos( \pi \theta / \theta_c) + 1\right) \left(\cos(\pi \phi / \phi_c) + 1 \right)
\label{eq_potential}
\end{eqnarray}
with $\LJ{p}$ a generalized truncated and shifted Lennard-Jones function:
\begin{align}
\LJ{p}(x,x_\mathrm{c},\sigma) \equiv &
      4 \left( \left(\frac{x}{\sigma}\right)^{-p} - \left(\frac{x}{\sigma}\right)^{-p/2}  \right. \nonumber \\
        & -\left. \left(\frac{x_\text{c}}{\sigma}\right)^{-p} + \left(\frac{x_\text{c}}{\sigma}\right)^{-p/2} \right) \Theta(x-x_\text{c})
\label{eq_LJ}
\end{align}
In Eq. \eqref{eq_potential} the index $b$ sums over pairs of complementary bond vectors, $\Theta(x)$ is the Heaviside step function and  $R_{ij}$ is the subunit center-to-center distance.

The nanoparticle  is modeled as a spherical excluded volume with a Lennard-Jones potential whose argument is shifted so that the minimum occurs when subunits are on the surface:
\begin{align}
U_\text{S}(r) &= \eS \LJ{12}(r_\mathrm{eff}, \rc, \sigma) \nonumber \\
r_{\mathrm{eff}} &\equiv r - R_\text{S} + 0.5\sigma
\label{eq_sphereU}
\end{align}
 with $r$ as the nanoparticle-subunit center-to-center distance, $R_\text{S}$ as the radius of the nanoparticle, and $\eS$ as a tuneable subunit-nanoparticle interaction strength.  This potential qualitatively represents subunit electrostatic interactions with functionalized gold nanoparticles\cite{Tsvetkova2012, Sun2007, Huang2007, Dixit2006, Daniel2010, Chen2005, Chen2006}.

{\bf Units and parameter values.}  Lengths have units of $\sigma$, the subunit diameter, energies have units of $\kt$, and times have units of $t_0 = \sigma^2/D$, where $D$ is the subunit diffusion constant.
The box side length is $18\sigma$, the nanoparticle radius is $R_\text{S}=0.9\sigma$, grand canonical particle exchanges occur at least $9\sigma$ from the nanoparticle surface, and the reservoir subunit concentration is $c_0=0.005 \sigma^{-3}$ (defined as $c_0=N \sigma^3/L^3$ with $N$ the number of subunits).   The attractor cutoff values are $\rc=2.5\sigma$, $\tc=1$ and $\pc=\pi$.

\subsection{Triangles model}
\label{sec:trianglesAppendix}
\begin{figure}[hbt]
  \begin{center}
  \includegraphics[width=0.6\columnwidth]{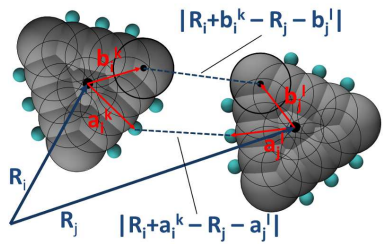}
  \caption{Triangles model subunit geometry with distances marked.  Excluders are gray, subunit attractors are teal, and the black circles depict the overlapping arrangement of excluders (polymer attractors not shown).  Note that the position of the $k^\mathrm{th}$ attractor on subunit $i$ is given by $\bm{r}_i^k \equiv \bm{R}_i + \bm{a}_i^k$, and the position of the $l^\mathrm{th}$ excluder on subunit $j$ is given by $\bm{r}_j^l \equiv \bm{R}_j + \bm{b}_j^l$ (as used in Eqs \eqref{eq_Ucc} and \eqref{eq_Ucp}).}
  \label{fig_triangles_geometry}
  \end{center}
\end{figure}

In this model taken from Ref. \cite{Elrad2010}, the minimum energy structure is a complete capsid of 20 subunits assembled around a bead-spring polymer.  The truncated pyramidal subunits are composed of two layers of excluders, which enforce excluded volume interactions, and two layers of capsomer attractors on the edges, which mimic hydrophobic and electrostatic attractions (see Fig. \ref{fig_triangles_geometry}).  The attractive interaction between subunits is maximized when the capsomer attractors are perfectly overlapping.  In this situation, excluders on either side of the interface are separated by the cut off distance of their potential.

The polymeric cargo is  a freely-jointed chain of spherical monomers, which have an excluded volume that represents steric interactions and screened electrostatic repulsions.  The model represents the polymer in good solvent, which behaves as a self-avoiding random walk with radius of gyration $R_\text{g} = 0.21N_\text{p}^{3/5}\sigma_\text{b}$; $N_\text{p}$ is the number of monomers and $\sigma_\text{b}$ is the monomer diameter.  The polymer has short-range interactions with the polymer attractors on the bottom of each subunit, which represent screened electrostatic interactions.  All interactions in this model can be decomposed into pairwise potentials.

\textbf{Capsomer-Capsomer Interactions.}
Capsomer subunit interactions decompose into pairwise interactions between their constituent building blocks -- the excluders and attractors.  Excluders in a subunit $i$ experience a truncated Lennard-Jones-like potential when within a cutoff radius $r_\text{c}$ of excluders in subunit $j \neq i$ (i.e. subunits excluders repel each other when overlapping).  Attractors in subunit $i$ experience an attractive Lennard-Jones-like potential with \emph{commensurate attractors} from another subunit $j \neq i$.  Commensurate attractors are defined to be the attractors which would overlap in a complete capsid \cite{Elrad2010}:
\begin{align}
	U_\text{cc}(i,j) &= \sum_{k,l\ \in \mathrm{\ excluders}} \LJ{8} \left( \left|\bm{r}_i^k - \bm{r}_j^l\right|,2^{1/4}\sigma_b,\sigma_b \right) \nonumber \\
	&  + \sum_{k,l\ \in \mathrm{\ attractors}} \ecc \LJ{4} \left( \left|\bm{r}_i^k - \bm{r}_j^l\right| - 2^{1/2}\sigma_a, 4\sigma_a, \sigma_a \right)
\label{eq_Ucc}
\end{align}
where the first sum is over all pairs of excluders between subunits $i$ and $j$ and the second sum is over all commensurate attractor pairs between subunits $i$ and $j$, with $\ecc$ as an adjustable parameter that tunes the attraction strength.  The coordinates of the $k^\mathrm{th}$ excluder/attractor are given by $\bm{r}_i^k$ and $\LJ{4}$ is given by Eq.~\eqref{eq_LJ}.

\textbf{Capsomer-Polymer Interactions.}
The capsomer-polymer attraction is similar to the capsomer-capsomer interaction, with an attractive potential that is minimized when polymer beads overlap subunit attractors and a repulsive potential to account for excluded volume. For a capsid subunit $i$ and polymer subunit $j$, the potential is
\begin{align}
\label{eq_Ucp}
	&U_\text{cp}(i,j) = \sum_{k\ \in \mathrm{\ excluders}} \LJ{8} \left( \left|\bm{r}_i^k - \bm{r}_j \right|,2^{1/4}\sigma_\text{bp},\sigma_\text{bp} \right) \nonumber \\
	& + \sum_{k\ \in \mathrm{\ poly\ attractors}} \xi_k \ecp \LJ{8} \left( \left|\bm{r}_i^k - \bm{r}_j\right| +2^{1/4}\sigma_p, 4\sigma_\text{p}, \sigma_\text{p} \right) \\
	& \sigma_\text{bp} \equiv \frac{1}{2} \left( \sigma_\text{b} + \sigma_\text{p} \right) \nonumber
\end{align}
with the first sum over all subunit excluders and the second sum over all polymer attractors on the subunit.  $\ecp$ parameterizes the attraction strength for each attractor.  $\sigma_\text{p}$ is the diameter of a polymer bead and is set to $0.4\sigma_\text{b}$.  $\xi_k$ is a factor that decreases the interaction strength for the outmost three polymer attractors on the subunit ($\xi_k = \frac{1}{2}$) to compensate for these sites overlapping in a complete capsid.  $\xi_k$ is set to one for all other attractors.

\textbf{Polymer-Polymer Interactions.}
The polymer potential includes a `bonded' interaction between monomers that are nearest neighbors along the chain and a nonbonded, excluded volume interaction with all other polymer beads.  For bead coordinates $\bm{r}_i$ and $\bm{r}_j$ ($i \neq j$)
\begin{align}
&U_\text{pp}{}(r_{ij} \equiv |\bm{r}_{i} - \bm{r}_{j}|) \nonumber \\
&=
\left\{
	\begin{array}{ll}
	\LJ{8}(r_{ij}, 2^{1/4}\sigma_\text{p}, \sigma_\text{p})	& : r_{ij} < 2^{1/4} \sigma_\text{p} \\
	\LJ{8}(2^{5/4}\sigma_\text{p} - r_{ij}, 2^{5/4}\sigma_\text{p}, \sigma_\text{p})	& : r_{ij} > 2^{1/4} \sigma_\text{p} \ \\
	&	\& \ \{i,j\}\ \mathrm{bonded} \\
	0	& : \mathrm{otherwise} \\
	\end{array}
\right.
\label{eq_upp}
\end{align}

{\bf Units and parameter values.} The capsomer subunits have anisotropic translational and rotational diffusion constants calculated using Hydrosub7.C~\cite{Torre2002,Elrad2010}. Lengths are scaled by $\sigma_\text{b}$ and times are scaled by  $t_0$, which is the Brownian time for a sphere with diameter $\sigma_\text{b}$. The box side length is $40\sigma_\text{b}$, grand canonical moves are performed at least $10\sigma_\text{b}$ from the center of the polymer, the reservoir concentration is $0.00047 \sigma_\text{b}^{-3}$, and the polymer length is 200. The attractor and polymer bead diamaters are respectively $\sigma_\text{a}=0.2\sigma_\text{b}$ and $\sigma_\text{b}=0.4\sigma_\text{b}$.

\section{Collective coordinates}
\label{appendix_graphIso}

\begin{figure}[hbt]
  \begin{center}
  \includegraphics[width=0.6\columnwidth]{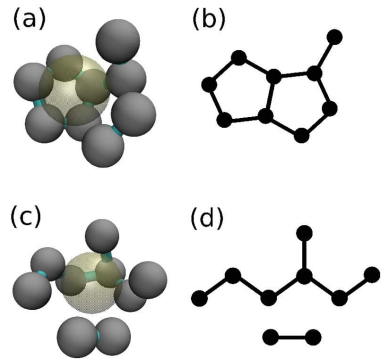}
  \caption{{\bf (a)} and {\bf (c)} show clusters of patchy spheres growing on the nanoparticle during assembly while {\bf (b)} and {\bf (d)} show their respective graph representations.  Attractors are not shown, but bonds (strong attractions) between subunits are depicted as teal cylinders.  In this work, only the largest cluster was considered when constructing the MSM, which would exclude the lower dimer in (c),(d) from consideration.}
  \label{fig-graphIso-explanation}
  \end{center}
\end{figure}

\textbf{Graph Isomorphism.}
 Here we describe a general approach to state decomposition for self-assembly reactions, which represents the bonding network of a growing cluster as an undirected graph (Fig.~\ref{fig-graphIso-explanation}), with subunits as the nodes and strong interactions, which we denote as `bonds', as the edges (see Fig. \ref{fig-graphIso-explanation}). Structures which differ only in the indices that label their subunits or fluctuations of bond geometries have the same graphs; to account for subunit index changes, isomorphic graphs were identified using the algorithm in the Boost C++ libraries \cite{Fortin1996,Reingold1977}, although faster algorithms are available (e.g. \cite{McKay1981}). We only applied this order parameter to class I subunits (i.e. the largest growing cluster) as described in section \ref{sec_build_msm}, but it can be generalized to class II subunits by considering subunit interactions with the scaffold.

\textbf{Alternative structure-based order parameters.}
Through \emph{a priori} knowledge of the system, it is possible to devise state decompositions which result in fewer states. For example, subunits attached to a cluster by only a single interaction are transient under typical assembly conditions \cite{Hagan2006}.  Graphs can therefore be simplified by pruning nodes which are connected to the rest of the graph by only a single edge. Further simplifications can be made by only recording features of the graph.  For systems in which single bonds are relatively unstable, clusters tend to grow by sequential completion of polygons \cite{Hagan2011}.  One can then record the set of  all cycles (which correspond to complete polygons) in the graph, or one can account only for the number of subunits $n$ and number of bonds $n_\text{b}$ within a growing cluster.  These two quantities can be combined into a single coordinate as
\begin{equation}
\gamma = a (n_{\mathrm{b}} - n) + n_{\mathrm{b}}
\label{eq_assembly_op}
\end{equation}
with a parameter $a$ that prevents degeneracy among typical structures by separating structures with different numbers of complete polygons. We used $a{=}5$ or $a{=}10$.

\section{Simulations and protocol for estimating the transition matrix}
\label{sec:protocol}
This section describes the ratcheting procedure that we used to estimate the elements of the transition matrix.  We began by performing $\ns$ simulations (with $\ns{=}50-100$), each started from initial configurations in which subunits had random positions and orientations (except that subunits were not allowed to overlap). Each simulation was run for a time $\ts$, and snapshots were saved regularly to a database.  Saved configurations were then classified into states.  This classification could be based on any of the state decomposition approaches described in this article; we used the coordinate defined in Eq.~\eqref{eq_assembly_op}. A second iteration was then started, in which $\ns$ new simulations of length $\ts$ were performed with snapshots regularly saved to the database.  To efficiently estimate important elements of the transition matrix, initial configurations were preferentially chosen from states from which fewer simulations had already been initialized. Iterations were then repeated until the MSM converged (Fig.~\ref{fig_convergence}).  This procedure is useful when no \emph{a priori} knowledge is available about the system, since it both focuses sampling on poorly sampled states and identifies new states that arise through the natural dynamics of the system. However, once enough states have been gathered to construct an MSM, more sophisticated adaptive sampling is possible \cite{Bowman2010}.

Optimal values of the parameters $\ns$ and $\ts$ depend on both the system being simulated and the available computational resources and thus need to be chosen through trial and error. The simulation time $\ts$ must be longer than the lag time $\tau$, but should not be too long in order to efficiently ratchet the system over free energy barriers. We found that a $\ts$ of around $20\tau - 100\tau$ usually worked well. The total number of simulations required to generate a converged MSM varied depending on the parameters, but we found that about 1,000 simulations were sufficient for most parameter sets. We used a modified version of MSMBuilder \cite{Bowman2009a} to construct the transition matrix and to calculate its eigenvalues and eigenvectors.  For most parameter sets, it was sufficient to calculate only the top 20 eigenvectors.

For the models that we consider here, dissociation of a subunit from a complete capsid occurs only on timescales long compared to the assembly time for most parameter sets. Thus, well-formed target states are effectively absorbing states on simulated timescales (see Refs. \cite{Hagan2010,Zlotnick2007,Hagan2014} for discussion); however, on rare occasions, capsids would form in strained configurations and eventually dissociate.  When building the transition matrix, it was important to acquire sufficient statistics from short trajectories in the target state to balance the rare dissociations from strained target states. This need could be avoided by either more stringently defining the target state or by manually defining the target state as an absorbing state during transition matrix construction.  Note that if one were interested in capturing the much longer time scale associated with dissociation of a well-formed target structure, additional coordinates describing fluctuations of the subunit-subunit bonds would be required to efficiently sample this transition.



%

\end{document}